\documentclass[aps,twocolumn, showpacs]{revtex4}
\usepackage{graphicx}
\usepackage{dcolumn}

\begin{document}


\title{Tuning the ferromagnetic properties of hydrogenated GaMnAs}

\author{L. Thevenard}
\author{L. Largeau}
\author{O. Mauguin}
\author{A. Lema\^{\i}tre}
\email[]{Aristide.Lemaitre@lpn.cnrs.fr}
\affiliation{CNRS / LPN- Laboratoire de
Photonique et de Nanostructures, Route de Nozay,
91460 Marcoussis, France}
\author{B. Theys}
\affiliation{Groupe de Physique des Solides, Universit\'{e}s Paris
VI et VII, 4, place Jussieu, 75252 Paris Cedex 05, France}


\date{\today}

\begin{abstract}
Hydrogenation and post-hydrogenation annealings have been used as a very efficient tool to
tune the hole density over a wide range, at fixed magnetic moment
concentration in thin GaMnAs layers. Reduction of the hole density resulted 
in strong modifications of their ferromagnetic properties. In particular, we
observed in magneto-transport experiments the decrease of the Curie temperature,
along with modifications
of the magnetic anisotropy, a behavior predicated by the mean-field theory.
\end{abstract}


\maketitle

Carrier-induced ferromagnetism in diluted magnetic semiconductors
is mediated by the exchange interaction between the delocalized carriers
and the magnetic ions.\cite{Matsukura02} In GaMnAs, due to the Mn
acceptor character, it is thus expected, and it has been predicted\cite{Dietl01a,Abolfath01}
and demonstrated,\cite{Shen97,Wang04} that the ferromagnetic properties,
as the Curie temperature ($T_{C}$), strongly depend on the hole density
($p$) and the precise shape of the valence band. In particular, \cite{Dietl01a,Abolfath01}
have predicted a strong influence of the carrier density on the magnetic
anisotropy, leading to the re-orientation of the easy axis at low
$p$ for a given epitaxial strain.

A better understanding of the influence of the carrier density on
the ferromagnetic properties would require data obtained from experiments
in which the density can be tuned over a large range for a given Mn
concentration. However, in GaMnAs, $p$ is fixed, a few 10$^{20}$~cm$^{-3}$,
by the number of incorporated Mn ions, since they both act as magnetic
moments and acceptors, even so donor defects have been shown to compensate
partially the substitutional Mn atoms.\cite{Yu03} It is therefore
difficult to separate both contributions, hole density and Mn concentration,
to the ferromagnetic phase. Several techniques have been reported
in GaMnAs such as co-doping with Sn donors\cite{Satoh01} or post-growth
annealing which allow the control of $p$ for a rather fixed Mn concentration.\cite{Sawicki04}
However these techniques are limited either to a single hole density
or to a rather narrow range of densities (from $5\times10^{19}$ to
$1\times10^{20}$~cm$^{-3}$ in the case of Ref. \onlinecite{Sawicki04}).

Having shown recently\cite{Bouanani03,Lemaitre05} that hydrogen effectively
neutralizes Mn in GaMnAs, we have used hydrogenation as a very efficient
way to strongly reduce $p$ by neutralizing the Mn atoms, while keeping
a constant number of magnetic moments. This technique was also recently
proposed and experimentally demonstrated by an independent study.\cite{Brandt04,Goennenwein04}
Here, using subsequent thermal annealing, we were able to remove gradually
H atoms from the neutralized GaMnAs layer and to recover the hole
density and so the ferromagnetic phase. This technique allows the
fine tuning of $p$ over a large range. In this letter we report on
the dependence of GaMnAs magnetic properties on $p$, adjusted by
means of hydrogenation. The results presented here were obtained from
magneto-transport experiments.

Hydrogen has been widely used in p-doped GaAs to neutralize acceptors
(A).\cite{Chevallier91} Hydrogenation is performed by exposing the
sample to H or deuterium plasmas. Monoatomic H atoms diffuse into
the doped layer and form electrically inactive H-A complexes. In GaMnAs,
hydrogen incorporation and Mn neutralization were evidenced by secondary
ion mass spectroscopy and by local vibrational mode spectroscopy respectively.\cite{Bouanani03,Brandt04}
In this study, the sample consists of a 50~nm thick Ga$_{1-x}$Mn$_{x}$As
layer grown on (001) undoped GaAs substrates at around 260$^{\circ}$C.
The manganese concentration is estimated to be $x\simeq6-7\%$. Prior
to hydrogenation, the sample was annealed under nitrogen atmosphere
for 1~hour at 250$^{\circ}$C to strongly reduce and homogenize the concentration of highly
mobile interstitial Mn atoms, insuring that the subsequent modifications
of the magnetization were due to the H neutralization only (the decrease of interstitial atoms was checked using the technique presented in ref.~\onlinecite{Glas04}). The sample
was then hydrogenated during 3~h at 130$^{\circ}$C using a direct
exposure radio-frequency H plasma (13.56~MHz, 80~mW.cm$^{-2}$,
1~mbar). Also, the Hall bars were processed before hydrogenation
to avoid the neutralization of the GaMnAs regions underneath the Ti/Au
layers insuring good ohmicity of the contacts. Optimization of temperature
and duration was done by monitoring the change of the resistance of
the device before and after hydrogenation. After hydrogenation, values
larger than 40~M$\Omega$ were obtained similar to semi-insulating
GaAs substrate resistance, giving an upper limit for $p$ of $6\times10^{16}$~cm$^{-3}$
assuming a mobility of 5~cm$^{2}$/V/s (from Ref.~\onlinecite{Goennenwein04})
This indicates that H atoms are highly mobile in the GaMnAs layer
and efficiently passivate Mn atoms, even at such low temperature.
Higher hydrogenation temperatures, above 170$^{\circ}$C, resulted,
for the same exposure times, to lower device resistance, so lower
H incorporation. Last, a reference sample was obtained by covering
it with a GaAs substrate during exposure to the plasma, to prevent
hydrogenation.

\begin{figure}
\includegraphics[width=8.5cm]{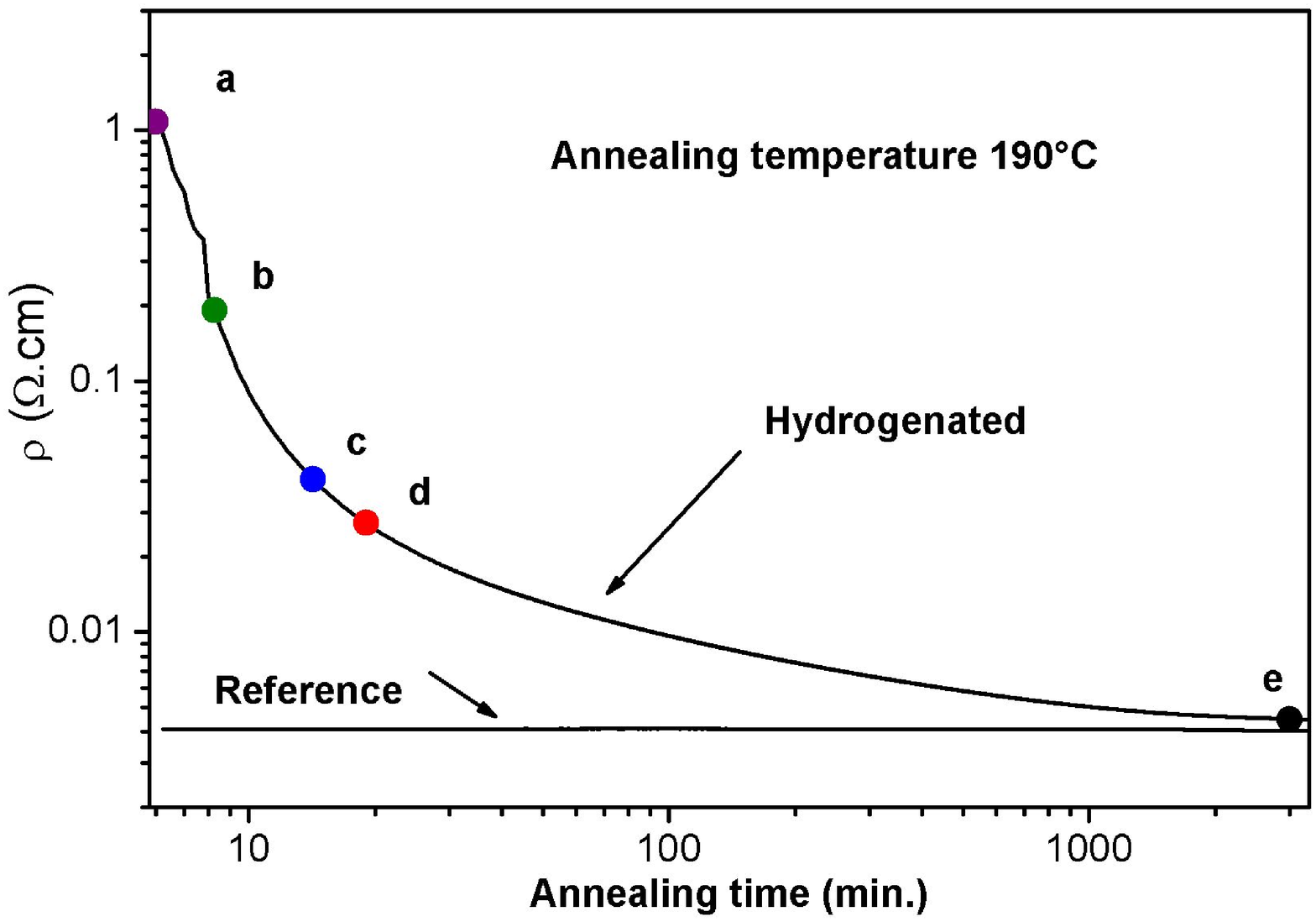}
\caption{\label{Recuit} (Color online). Experimental temporal evolution of the resistivity
upon annealing of the hydrogenated  a, b, c, d and e and reference pieces, monitored during the annealing at 190$^{\circ}$C. The annealing
times and resistivities corresponding to the hydrogenated samples
reported in this study are represented by the dots.}
\end{figure}

The formation of Mn-H complexes upon hydrogenation resulted in a strong
increase of the resistivity, due to the large reduction of the hole
density. Thermal annealing has been used in hydrogenated p-doped GaAs to break
the A-H complexes, to remove the H atoms from the layer, resulting
in the re-increase of the hole density. In GaMnAs, however, the maximum
annealing temperature is limited by the tendency of GaMnAs to deteriorate
at temperatures above 300$^{\circ}$C. Nevertheless our experiment indicated
that temperatures as low as 190$^{\circ}$C are sufficient to get
large removal of H atoms within a few hours. Figure~\ref{Recuit}
shows the resistivity evolution with annealing time for both hydrogenated
and reference samples, monitored \textit{in situ} during the annealing
under nitrogen atmosphere (annealings under vacuum gave similar results).
The reference sample exhibits almost no change as Mn interstitial
atoms play a minor role, since they were stabilized during the pre-annealing,
giving a first indication that the magnetic moment concentration remains
constant during the post-hydrogenation annealing. On the opposite,
the hydrogenated one shows a very large decrease of the resistivity,
by more than two orders of magnitude. After $\sim$3000~min, the
resistivity reaches the value measured for the reference sample within
10~\% (the usual deviation for these Hall bar devices), indicating
i) that most of the H atoms left the GaMnAs layer and ii) that the
hole density recovers its original value (an increase of the mobility
is quite improbable).

Post-hydrogenation annealings offer a very practical and non-destructive
way to precisely tune the hole density, over a large range by adjusting
the annealing time. At this point, we would like to underline the
difficulty to determine accurately the hole density, because of the
anomalous Hall effect contribution. So at the present, we do not have
such information. High magnetic fields measurements\cite{Edmonds02}
or Raman spectroscopy\cite{Limmer02} may solve this issue. However,
this determination could be further hampered at low densities, when
crossing the metal-insulator transition.

\begin{figure}
\includegraphics[width=8.5cm]{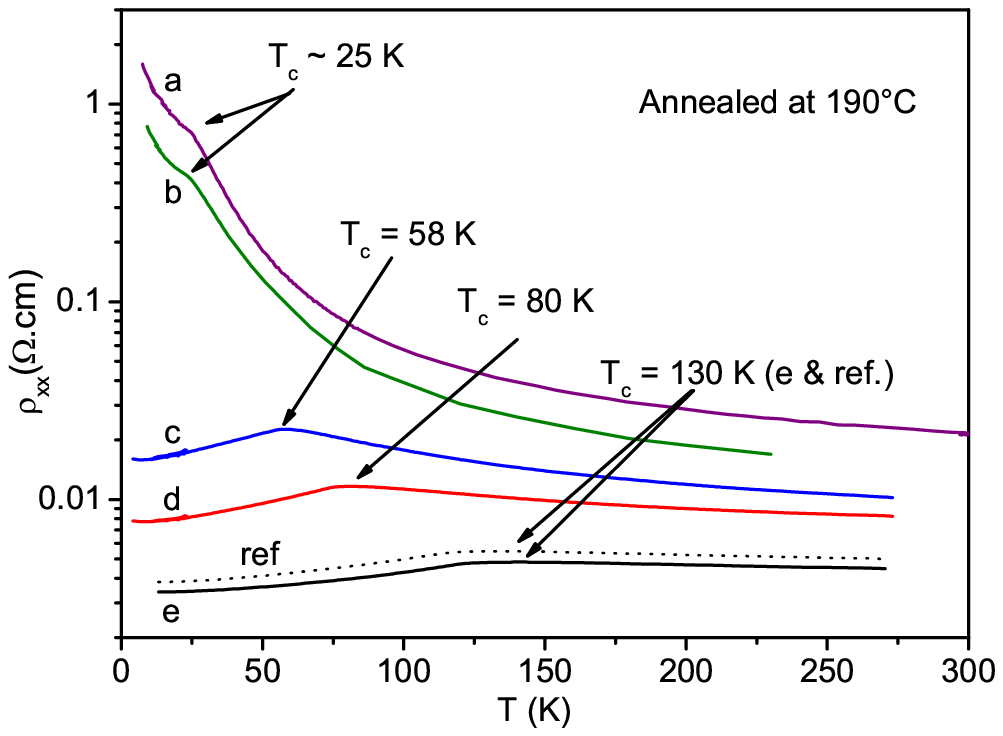}
\caption{\label{Resistivity} (Color online). Longitudinal resistivity ($\rho_{xx}$) vs.
temperature for post-hydrogenation annealed pieces a, b, c, d, e and
the reference sample (dotted line). $T_{C}$ is estimated from the resistivity maximum.}
\end{figure}

Change of the carrier density induced large changes in the magnetic
properties of GaMnAs. Here, we investigated them through the temperature
dependence of the sheet resistivity and the anomalous Hall effect,
which both require sufficient carrier density. In particular, this
type of experiments does not give access to the regime where the ferromagnetism
disappears,\cite{Brandt04,Lemaitre05} since it occurs at vanishing
$p$, where transport measurements are not applicable. Magnetic measurements,
which require great care to avoid parasitic signals, especially for
such thin layers, will be presented elsewhere. Several pieces of the
same sample, respectively labeled a, b, c, d and e, were annealed
for different post-hydrogenation annealing times, 11, 13, 19, 24 and
3000~min, respectively, thus corresponding to different hole densities.
The temperature dependence of the resistivity is presented in Fig.~\ref{Resistivity}
along with the one measured for the reference sample. For high $p$
(annealings c, d, e and the reference sample), the curves show a clear
maximum, followed by a moderate decrease at lower temperature, suggesting
a metallic behavior. At lower $p$ (annealings a and b), the curves
exhibit a bump as the resistivity strongly increases at low temperature,
indicating an insulating character. These peaks or bumps correspond
approximatively to $T_{C}$.\cite{Matsukura02} As expected, post-hydrogenation
annealings resulted in increasing $T_{C}$ with increasing annealing
time, since the hole density is progressively restored when H atoms
leave the GaMnAs layer. After 11 and 13~min, a bump is seen around
25~K, while longer annealings yield, respectively $T_{C}=$ 58 and
80~K after 19 and 24~min. After 3000~min annealing, the Curie temperature
is identical to the one measured for the reference sample, $T_{C}=130$~K,
indicating complete recovery of the ferromagnetic phase observed prior
to hydrogenation.

\begin{figure}
\includegraphics[width=8.5cm]{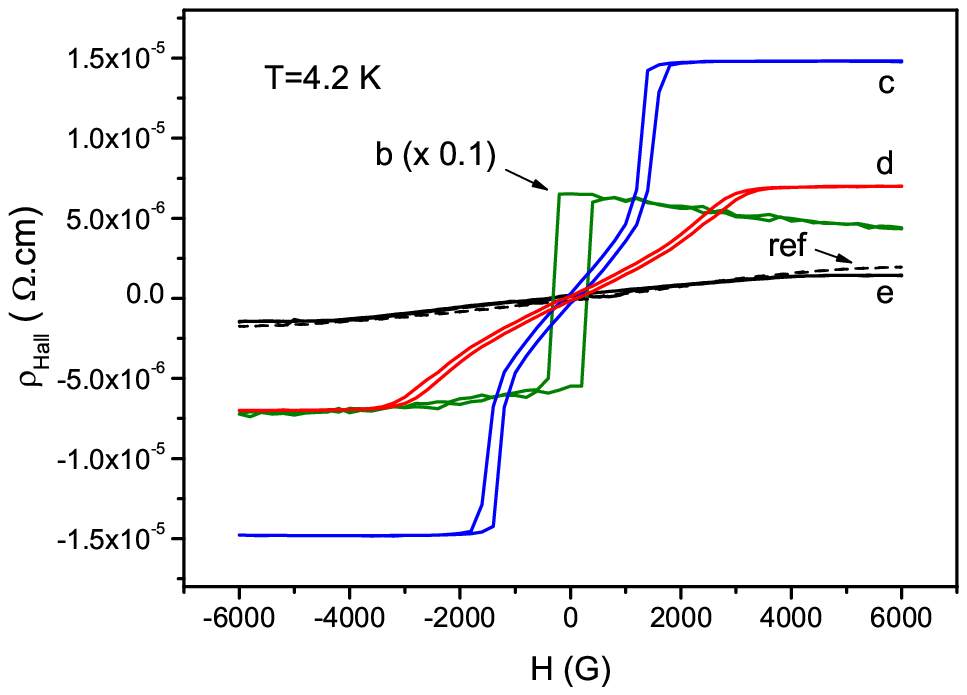}
FIGURE \ref{Hysteresis}
\caption{\label{Hysteresis} (Color online). Hall resistivity vs magnetic field measured for
samples b to e and ref (dotted line). Curves c and d have been corrected for a small planar Hall effect contribution.}
\end{figure}

We now turn to the magnetic anisotropy in these samples, since it
is expected to strongly depend on $p$. Calculations within the mean-field
theory\cite{Abolfath01,Dietl01a} have indeed predicted that the magnetic
anisotropy in GaMnAs mostly stems from the anisotropy of the carrier-mediated
exchange interaction. In particular, the in-plane easy axis, usually
observed in GaMnAs layers grown on GaAs, arise from the compressive
epitaxial strain, whereas layers under tensile strains generally exhibit
an out-of-plane magnetic anisotropy. However recent experiments\cite{Takamura02,Sawicki04}
have shown that for low values of $p$ and temperature a reorientation
of the easy axis, from in-plane to out-of-plane in theses cases, occurs,
a behavior consistent with mean-field predictions,\cite{Dietl01a,Abolfath01}
and related to the filling of the different valence sub-bands. Post-hydrogenation
annealings offer a unique opportunity to investigate the modification
of the anisotropy upon $p$ at fixed magnetic moment concentration.
Fig.~\ref{Hysteresis} shows the Hall resistivity $\rho_{Hall}$,
measured at 4~K, for the annealed and reference samples. In GaMnAs,
the Hall effect is dominated by the anomalous Hall effect which is
proportional to the sample magnetization perpendicular to the growth
axis.\cite{Matsukura02} Note that we did not plot $\rho_{Hall}/\rho_{xx}^{\gamma}$
with $\gamma=1$ as usually found in the literature as the $p$ dependence
of the different spin scattering mechanisms is rather complex.\cite{Jungwirth02}
However, this choice has little impact on the shape of the curves
in Fig.~\ref{Hysteresis}. Sample~e, completely depassivated, exhibits,
as expected, an almost identical behavior as the reference sample,
namely a progressive increase of the Hall resistivity and a saturation
at around 0.4~T, which corresponds to the progressive alignment of
the magnetization along the hard axis, in this case parallel to the
growth axis {[001]}. Sample b shows an opposite behavior since a
hysteresis cycle is seen, indicating that {[001]} is now an easy
axis. Samples c and d exhibit an intermediate regime. Although no
clear hysteresis cycle is observed, sample c magnetization first increases
progressively and then rapidly saturates at $\approx$~0.15~T, suggesting
that {[001]} is a less easy axis. Sample d magnetization evolves
quite smoothly and saturates at 0.3~T, indicating that {[001]}
is no longer an easy axis. A smaller saturation field however denotes
a weaker magnetic anisotropy compared to sample e. Such observations
confirm that the magnetic anisotropy is strongly dependent on the
hole density. In particular, as predicted by mean-field theories,\cite{Dietl01a,Abolfath01}
the anisotropy field is reduced when the hole density is lowered (corresponding
to shorter post-hydrogenation annealing time), even leading at lower
$p$ to an easy axis parallel to the growth axis {[001]}.

In conclusion, we have devised a simple technique based on post-hydrogenation
annealings of GaMnAs layers to finely tune the hole density, over
a large range, at fixed magnetic moment concentration. We have shown,
through magneto-transport measurements its ability to investigate
in detail the mechanisms responsible for the ferromagnetic phase in
this compound, especially the strong dependence of the magnetic anisotropy
on the hole density, a specific property of these systems. It should
be of great help to assess the validity of the theoretical models,
in particular if going beyond the mean-field theory is required or
not. 

\begin{acknowledgments}
This work has been supported by the R\'{e}gion \^{I}le de France,
the Conseil G\'{e}n\'{e}ral de l'Essone and through the Actions
Concert\'{e}es Nanostructures DECORESS and Incitative BOITQUAN. 
\end{acknowledgments}
\pagebreak

\end{document}